\documentclass[manuscript]{aastex}

\usepackage{natbib}
\usepackage{epsfig}
\usepackage{emulateapj5}

\slugcomment{Accepted by the Astrophysical Journal, to appear Oct 2004.}

\newcommand{\be}{\begin{equation}}
\newcommand{\ee}{\end{equation}}
\newcommand{\ba}{\begin{eqnarray}}
\newcommand{\ea}{\end{eqnarray}}

\begin{document}

\title{The Effect of Neutron Star Gravitational-Binding Energy on 
Gravitational-Radiation-Driven Mass-Transfer Binaries}

\author{Evelyne Al\'ecian\altaffilmark{1} and Sharon M. Morsink}
\affil{Theoretical Physics Institute, Department of Physics,
University of Alberta, Edmonton AB, Canada, T6G~2J1}

\altaffiltext{1}{Present Address: Universit\'{e} Paris VII, 2 place Jussieu, 
75251 Paris cedex 05, France}

\begin{abstract}
In a relativistic model of a neutron star, the star's mass is 
less than the mass of the individual component baryons. This is
due to the fact that the star's negative binding
energy makes a contribution to the star's total energy
and its mass. A consequence of this relativistic mass deficit
is that a neutron star that is accreting matter increases
its mass at a rate which is slower than the mass of a baryon
times the rate that baryons are accreted. This difference
in the rate of change of the masses has a simple relation
with the star's gravitational redshift.
We show that this
effect has the potential to be observed in binaries where
the mass transfer is driven by angular momentum losses from
the gravitational radiation emitted by the binary motion,
if the physics of the donor star is well-enough understood.
\end{abstract}


\section{Introduction}

In the general theory of relativity, all forms of stress and
energy act as a source for the gravitational field. As a result,
the mass of a compact star is the total energy of the star
divided by $c^2$. As the total energy of a star includes 
both the rest-mass energy and the
negative  binding energy, the mass of the star
is smaller than the sum of the masses of its component 
particles. 
This mass deficit effect is roughly proportional to 
the star's compactness, so it is most important for neutron
stars. For neutron stars near the upper mass limit, the
mass deficit can be as large as 25\% of the star's mass 
(see for example, \citet{CST94}), depending on the equation
of state.

A consequence of the mass deficit is that when a
neutron star accretes matter, its mass changes at a rate
that differs from the rate that baryons are accreted times
the mass of a baryon. This 
mass deficit effect is then potentially observable through
changes in the orbital period of the binary caused by 
accretion, if the mass accretion rate is known. An important
class of binaries where this effect may be observable are
binaries in which the mass transfer is driven by angular
momentum losses due to the emission of gravitational
radiation caused by the orbital motion. In the standard
scenario for gravitational-radiation-driven mass transfer
\citep{KMG62,Fau71},
the mass deficit is neglected since the primaries 
are white dwarf stars. However, in binaries where
the primary is a neutron star, the effect is in principle
observable, and could lead to constraints on the neutron
star mass and equation of state. In this paper we will 
address this issue and discuss the conditions
where the relativistic mass
deficit's effects could be detected. The effect described
requires very precise timing of the binary's orbital period in 
order to be detected, so we will focus our attention on the
class of binaries containing an accreting millisecond X-ray pulsar. 
At present three binaries of this class have measured mass functions:
SAX J1808.4-3658 \citep{WvK98,CM98}, XTE J1751-305 \citep{Markwardt02}  and 
XTE J0929-314  \citep{Galloway02}. In addition two more pulsars, 
XTE J1807-294 \citep{Mark03} and XTE J1814-338 \citep{Stroh03} have recently
been discovered, although their mass functions are not yet measured.
 Precise timing of the X-ray pulse 
arrival times allows the detection of these binary systems' orbital periods
through the orbital Doppler effect. It is expected \citep{CM98}
that future observations will reveal an orbital period derivative.

The effect of changing a neutron star's binding energy in a binary system
has been examined by other authors in other situations. In the case 
of a binary system without mass-transfer, \cite{SS01} considered the
changes in a neutron star's binding energy due to magnetic dipole spin-down
and its effect on the orbital period. The magnitude of the effect examined
by \cite{SS01} is quite small, but may be measurable.
In the present paper we 
allow for mass transfer, which produces an effect that is much larger.
While accretion-induced changes in a neutron star's characteristics 
have been considered by \cite{Spy88}, his model was for a simple
Newtonian stellar model and did not consider changes in the orbital
period.

In section \ref{s:binding} we review the concept of the binding energy
of a star and discuss its value and its derivatives for relativistic models of neutron
stars. In section \ref{s:transfer} we review the scenario of gravitational-radiation-driven
mass transfer and include the mass deficit in the derivation. 
In section \ref{s:evol} we review the different theories of how the compact binaries of
interest may have evolved and discuss which theories would allow the desired measurements
of the relativistic effect. Finally
in section \ref{s:conclusions} we discuss the main theoretical uncertainties in this method.

\section{The Binding Energy of Compact Objects}
\label{s:binding}

If we define $N$ to be the total number of baryons in the star,
$m_B$ the mass of a baryon and $\varepsilon$ the binding
energy of the star, then the total mass, $M$, of the star is
\be
M = m_B N + \varepsilon/c^2.
\ee
The binding energy is the sum of the star's gravitational 
potential energy, the internal energy and the rotational
kinetic energy. If the star were to be disassembled into
its component baryons, the mass of the baryons would be
$M_0 = m_B N$. 
For a gravitationally bound system, 
the binding energy is negative, so $M$ is always smaller than
$M_0$. Using Newtonian arguments, the leading order
contribution to the binding energy
is $\varepsilon \sim -G M_0^2/R$, so that the mass is given by
\be
M = M_0 \left( 1 - \varepsilon_0\frac{GM_0}{c^2 R} \right)
\label{scale}
\ee
where $\varepsilon_0$ is a constant of order unity.
Equation (\ref{scale}) shows that the relativistic correction
to the mass of a star scales as the compactness $GM/(Rc^2)$.
Hence this correction is negligible for main sequence stars
and white dwarfs. However, for neutron stars, 
the compactness can be as large as $4/9$. 
The contribution of the binding energy to the mass can be 
significant for compact stars. 

It is useful to briefly review the concept of mass and binding energy by 
recalling their definitions for a spherical star in general
relativity (see box 23.1 of \citet{MTW} for
an elementary discussion). A zero temperature equation of state (EOS)
can be written as equations for $P(n)$ and $\epsilon(n)$ where $n$ is the 
number density of baryons, $P$ is the pressure and $\epsilon$ is the
energy density. The energy density includes both the matter's rest-mass
energy density and its internal energy density. Once the EOS is 
specified, the star's total mass (using Schwarzschild coordinates) is
\be
M = 4 \pi \int_0^R \epsilon r^2 dr.
\ee
This is the mass which is measured by Kepler's law when a satellite orbits
the star. For this reason, $M$ is often called the ``gravitational mass''.
The baryon mass, 
$M_0$,  of the star is given by the volume integral
of the baryon number density times $m_B$. Since the volume
element in Schwarzschild coordinates is $dV = 4 \pi r^2 \left(1 - 2m(r)/r\right)^{-1/2} dr$,
the baryon mass is
\be
M_0 = m_B N = 4 \pi m_B  \int_0^R n \left(1 - 2m(r)/r\right)^{-1/2}  r^2 dr.
\ee
Even in the case of an incompressible fluid, there is a difference between $M$ 
and $M_0$ due to the binding energy of the star. These definitions
can be extended to rotating stars (see for example, \citet{FIP86}). 
The result is a family of 
equilibrium stellar models with gravitational
masses that depend on the star's baryon number $N$ and its angular momentum $J$
\citep{Bar73}.

Consider the accretion of baryons with specific angular momentum
$\ell$ at the rate $\dot{N}$
onto a star.  Since the gravitational
mass is a function of the stars' baryon number and 
angular momentum,
the rate of increase of 
the gravitational mass is
\be
\dot{M} = \dot{N} \left( 
	\left(\frac{ \partial M}{\partial N}\right)_J +
	m_B \ell \left(\frac{ \partial M}{\partial J}\right)_N
	\right)
	.
\ee
The partial derivatives of the gravitational mass are
given by \citep{Bar73}
\be
\left(\frac{ \partial M}{\partial J}\right)_N = \frac{\Omega}{c^2},
\ee
where $\Omega$ is the star's angular velocity,
and 
\be
\left(\frac{ \partial M}{\partial N}\right)_J
	=  m_B {\Phi},
\ee
where $\Phi$ is the dimensionless chemical potential (or
injection energy) for bringing one baryon from infinity 
to the pole of the
rotating star. 
Using the metric of an axisymmetric rotating star,
given by \citet{FIP86}
\be
ds^2 = - e^{2\nu} dt^2 + e^{2\psi}\left(d\phi -\omega dt\right)^2
+ e^{2\mu}\left(dr^2 + r^2 d\theta^2\right),
\label{metric}
\ee
the chemical potential is given by
\be
\Phi = \left(e^\nu\right)_{p},
\ee
where the subscript ``p'' denotes that the function is 
evaluated at the star's spin pole.
In the limit of a nonrotating star,
\be
\lim_{\Omega \rightarrow 0} \Phi = \sqrt{ 1 - \frac{2GM}{Rc^2} }.
\ee
Typical values for the chemical potential for rotating neutron 
stars can be found in
the paper by \citet{FIP86}, however they define a quantity $\beta$
that is related to the chemical potential
by $\Phi = \sqrt{\beta}$. 
The chemical potential can be thought of as the specific energy 
required to bring a particle from infinity to the star. Since
a particle at rest at infinity has larger energy than
a particle at rest at the surface of the star (due to gravitational
redshift), energy is liberated during accretion. The quantity
$1-\Phi$ is the fraction of the energy liberated during accretion.

\begin{figure*}[t]
\epsscale{0.6}
\plotone{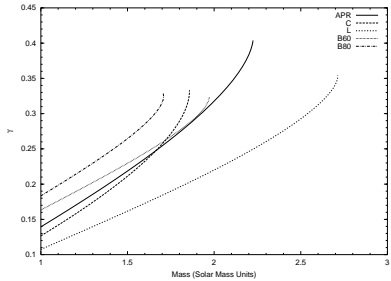}
\caption{Plots of the parameter $\gamma$ versus neutron
star mass for different models of nonrotating neutron stars.
\label{fig1}}
\end{figure*}

\begin{figure*}[b]
\epsscale{0.6}
\plotone{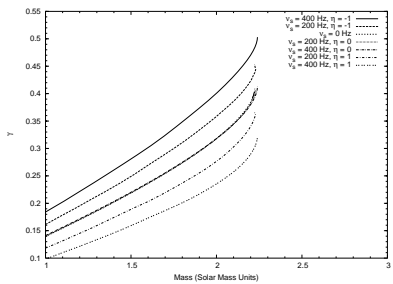}
\caption{Plots of the parameter $\gamma$ versus neutron
star mass for equation of state APR, 
for different stellar spin frequencies ($\nu_s$)
and torque parameters ($\eta$).\label{fig2}}
\end{figure*}

If we define a dimensionless angular velocity, $\bar{\Omega}$
\be
\bar{\Omega} = \Omega \sqrt{ \frac{R^3}{GM}}
\ee
and a dimensionless specific angular momentum, $\bar{\ell}$,
\be
\bar{\ell} = \frac{c \ell}{MG},
\ee
the rate of change of the star's gravitational mass due to accretion
is 
\be
\dot{M} = m_B \dot{N} \left( \Phi + \bar{\ell}\bar{\Omega} 
	\left( \frac{GM}{c^2R} \right)^{3/2}\right).
\ee
In general, not all of a particle's angular momentum will
be available to be added to the star's angular momentum. 
External torques, such as the star's magnetic field 
\citep{GL91} or
gravitational radiation \citep{Wag84,AKS99}
from nonaxisymmetric perturbations
may remove the angular momentum. It is useful then to 
introduce a torque parameter $\eta$ defined so that if the
baryon is in a circular orbit at radius $r$ around the star
when it is accreted, the specific angular momentum added
to the star is
\be
\bar{\ell} = \eta \bar{\ell}_K,
\ee
where $\bar{\ell}_K$ is the dimensionless specific 
angular momentum of a particle
in a circular Keplerian orbit. In the limit of a nonrotating star,
\be
\lim_{\Omega \rightarrow 0} \bar{\ell}_K = \frac{rc^2}{GM}
\sqrt{\frac{M}{rc^2/G-3M}}.
\ee
For a rotating star $\bar{\ell}_K$ is given by the 
quantity $cL/MG$, where $L$ is the specific angular momentum
defined in equation (7) of \citet{CST94}.
Since the rate that the star's mass increases will be close
to $m_B$ multiplied by the rate that baryons are accreted, we
will introduce a parameter $\gamma$, defined by
\be
\gamma = 1 - \frac{\dot{M}}{m_B \dot{N}}
	= 1 -  \Phi - \eta \bar{\ell}_K \bar{\Omega} 
	\left( \frac{GM}{c^2R} \right)^{3/2},
\label{gamma}
\ee
which is roughly 
proportional to the star's compactness. 
The parameter $\gamma$ corresponds to the fractional
rate of change of the relativistic mass deficit.
For a slowly rotating star, 
\be
\lim_{\Omega \rightarrow 0} \gamma = 1 -  \sqrt{ 1 - \frac{2GM}{Rc^2} }.
\ee
Recall that for a non-rotating star, the gravitational redshift,
$z$, is defined by
\be
z = \frac{1}{\sqrt{ 1 - \frac{2GM}{Rc^2} }} -1,
\ee
so in the limit of spherical symmetry, the parameter $\gamma$
is related to $z$ by $\gamma = z/(z+1)$.
For rotating stars, this relationship can be generalised by
defining $z_p$ as the gravitational redshift from the star's
pole. The parameter $\gamma$ is then related to the
polar redshift by 
$\gamma = z_p/(z_p+1) -
\eta \bar{\ell}_K\bar{\Omega}\left(GM/Rc^2\right)^{3/2}$.

In Figure \ref{fig1} plots of $\gamma$ versus gravitational
mass for non-rotating neutron stars are shown for a few
representative equations of state. We have chose three 
traditional neutron star equations of state (labeled C, L and APR)
and two simplified quark star (labeled B60 and B80)
equations of state. From the \citet{AB77} catalogue, we have
chosen equation of state C \citep{BJ74} 
which is moderately stiff and L \citep{PS75} which is very stiff
in order to show a typical range of stiffness for equations of
state. Equation of state APR \citep{APR98} makes use of 
data from modern
nucleon scattering experiments and also includes first order
special relativistic corrections. The quark star equations of
state are constructed in an approximation described in
\citet{Gle00} where the quarks' masses are set to zero
and the MIT bag model is used. The number after the letter ``B''
in the quark equations of state refer to the value of the 
bag constant, $B$, in units of $\hbox{MeV}/\hbox{fm}^3$.

The effects of rotation are shown in Figure \ref{fig2}, where
values of $\gamma$ for the APR equation of state are plotted
for different spin frequencies $\nu_s$  and torque values. 
The relativistic parameter $\gamma$ was computed by evaluating
equation (\ref{gamma}) after computing the value of the rotating 
star metric given in equation (\ref{metric}) using an accurate 2D
code\footnote[1]{The code is available at:
{\tt http://www.gravity.phys.uwm.edu/rns/ }} 
written by Stergioulas and described in \citet{SF95}.
This code is, in 
turn, based on the methods described by \cite{keh}
and \cite{CST94}. In the 
case where the accreted matter does not add any angular momentum
to the star ($\eta = 0$), it can be seen in Figure \ref{fig2}
that the function $\gamma$ is almost
independent of the star's spin frequency. The plots labeled
$\eta = 1$ correspond to stars where the magnetic field's effect
on the accretion disk is neglected so that the orbits extend down
to the marginally stable circular orbit and the accreted matter 
adds all of its angular momentum to the star. In this case, the
effect of the star's spin is to decrease $\gamma$ by a large
fraction over the nonrotating case. Since the value of $\ell_K$
has very little angular velocity dependence, it can be seen
that the effect of the star's angular velocity is to approximately
shift the curve downwards by an amount linear in the angular velocity.
If the torque parameter were negative, the curves
would be shifted vertically upwards by an amount linear in the
angular velocity. The magnetic disk models of \citet{RFS04} allow material to be accreted
without being propellered away from the star while decreasing the star's angular momentum.
In their models, the torque parameter can be negative and have a magnitude which is of
order unity. The curves labelled $\eta = -1$ correspond to this scenario and have 
an increased value of $\gamma$.

Simultaneous measurements of a neutron star's mass and the parameter $\gamma$ have
the potential to constrain the neutron star equation of state. Since $\gamma$ is
defined through the difference of $\dot{M}$ and $m_B \dot{N}$, accurate
measurements of these derivatives would be desirable. Clearly these are not 
 trivial measurements. In the highly idealized case of mass transfer 
driven by gravitational radiation, there is the potential to make this 
measurement if enough is known about the companion star. In the next section
we review the gravitational-radiation-driven mass-transfer scenario,
and discuss how $\gamma$ could be measured in an ideal situation. 
We will also discuss the uncertainties in the physics of the scenario
that will likely limit the ability to measure $\gamma$.

\section{Gravitational-Radiation-Driven Mass Transfer}
\label{s:transfer}

We now review the basic scenario for mass transfer driven
by angular momentum losses due to gravitational radiation. 
The gravitational-radiation-driven mass transfer scenario
was first developed \citep{KMG62,Fau71}
to explain the properties of white dwarf
binaries with short periods, however the physics also applies
to neutron star binaries.
In this section we will give a short review of the derivation of the
equations including the effects of both mass loss and
the neutron star's binding energy. The derivation of these equations
excluding the binding energy effect appears in many sources, such
as \citet{RJW82} and \citet{Ver93}.
 We will make use of the 
following notation: $M_1$ is the neutron star's gravitational
mass, while $M_2$ and $R_2$ are the mass and radius of the 
companion star. The mass ratio is denoted $q = M_2/M_1$.
For the ultra-compact systems which are most likely to be of interest,
the mass ratio is probably very small.

If there is no mass transfer between the stars, the orbital
separation will slowly shrink due to the emission of 
gravitational radiation. The binary system loses angular momentum
at the rate 
\be
\left(\frac{\dot{J}}{J}\right)_{GR} = -\frac{64 \pi}{5} q (1+q)^{-1/3} 
\left( \frac{2 \pi GM_1}{ c^3 P} \right)^{5/3} \frac{1}{P}
\label{Jdot-gr}
\ee
where $P$ is the orbital period. 
As the stars move closer together
the Roche lobe radius of the secondary star shrinks 
and eventually meets the surface of the secondary, at which 
point mass will begin to be transfered to the neutron star. 
Once mass transfer begins, the size of the secondary star
is assumed to be the same as the Roche lobe ($R_L$). In
the \citet{Pac71} approximation, the size of the Roche lobe
(and $R_2$) is given by
\be
R_2 = R_L = 0.46 a \left( \frac{q}{1+q} \right)^{1/3},
\ee
where $a$ is the distance between the stars. 
The time derivative of $R_2$ is 
\be
\frac{\dot{R_2}}{R_2} = \frac{\dot{a}}{a} 
	+ \frac13 \frac{\dot{q}}{q} \frac{1}{(1+q)}.
\label{adot}
\ee
The secondary loses mass at the rate $\dot{M_2}$. 
Since the secondary is assumed to be a non-relativistic
star (such as a white or brown dwarf), we do not need to 
make a distinction between the gravitational and baryon
masses of the secondary. As the secondary loses mass,
its radius is assumed to change as 
\be
\frac{\dot{R_2}}{R_2} = n_{ad} \frac{\dot{M_2}}{M_2} + \left(\frac{\dot{R}_2}{R_2}\right)_{th},
\label{Rdot}
\ee
where $n_{ad}$ describes the purely adiabatic changes to the star that depend only
on the mass loss, while $\left(\dot{R}_2/R_2\right)_{th}$ corresponds to changes in 
the star's structure due to thermal adjustments. This last term is of order $1/\tau_{KH}$,
where $\tau_{KH}$ is the Kelvin-Helmholtz timescale. While in most cases the thermal
timescale can't be neglected, in the case of a cool white dwarf the Kelvin-Helmholtz timescale
will be long enough that only terms depending on the mass loss rate need to be
included in equation (\ref{Rdot}).

In the most naive model of a degenerate star, $n_{ad} = -1/3$. However for very low
mass white dwarfs with finite temperature and an equation of state including 
Coulomb repulsion, the values of $n_{ad}$ range from -1/3 to 0 \citep{DB03},
depending on the mass, composition and temperature. If the mass and composition
are known, the uncertainty in temperature introduces an uncertainty in 
$n_{ad}$ of about 10\%.

For now, we will allow the most general mass transfer
and assume that only a fraction $\beta$ of the mass
lost from the secondary is captured by the neutron 
star. As a result, the baryon mass of the neutron 
star increases at the rate
\be
\dot{M}_{0,1} = - \beta\dot{M_2} .
\ee
However, the gravitational mass of the neutron star 
increases at a slower rate, given by 
\be
\dot{M_{1}} = - (1-\gamma)\beta\dot{M_2},
\label{Mdot}
\ee
where the parameter $\gamma$ (discussed in the previous
section) vanishes for newtonian stars
and is approximately equal to the neutron star's compactness.

Since we are allowing for non-conservative mass transfer, 
angular momentum will be lost.  Following the work of \citet{RJW82}, we parametrize
this angular momentum loss due to non-conservative mass transfer
with the parameter $\alpha$, so that
\be
\left(\dot{J}\right)_{\dot{M}} = \alpha (1-\beta) 
	\frac{2 \pi a^2}{P} \dot{M_2}.
	\label{Jdot-Mdot}
\ee
The parameter $\alpha$ encompasses the uncertainty in where the mass is lost.
If the mass loss occurs close to the centre of mass (say near the neutron star)
then $\alpha$ will be very small (of order $q^2$, see \citet{PRP02}), 
while if the mass loss occurs near the low
mass secondary, $\alpha$ will be close to unity. 

The liberation of energy during accretion due to the effects discussed
in this paper also introduce a mechanism for angular momentum loss. 
During the accretion process, the number of baryons and angular momentum
transfered to the star are conserved, while energy is liberated 
at the rate $\dot{E} = \beta \gamma \dot{M}_2 c^2$ in the rest frame
of the star. However, the star is orbiting the centre of mass at
a distance of $a_1 = q a/(1+q)$. Introducing a parameter $\delta$,
of order unity, the angular momentum loss due to the orbital 
motion of the neutron star as the energy is liberated is
\be
\left(\dot{J}\right)_\gamma = \delta \beta \gamma 
\frac{2 \pi a^2}{P} \left(\frac{q}{1+q}\right)^2 \dot{M}_2.
\label{Jdot-gamma}
\ee
In the systems of interest, the mass ratio $q$ is likely to be
very small, so this term will not be of much importance.

Another mechanism for angular momentum loss is magnetic braking,
where the donor star's stellar wind causes the donor to lose 
angular momentum through a coupling with the star's magnetic field.
If the binary system is tidally locked to 
the donor's spin, this causes the binary system to lose orbital
angular momentum \citep{VZ81}. We denote this mechanism through
a term $\dot{J}_B$ which is independent of the mass loss rate.
Estimates of the magnetic braking law depend on knowledge of
main sequence stellar winds and rotation rates. 
This type of mechanism is most likely to be of importance
if the donor was originally a main sequence star. The effects
of different parameterizations of $\dot{J}_B$ have been
explored by \citet{RVJ83}. 

Since the mass accreted onto the neutron star adds angular momentum to the star, this
too creates an effective loss of orbital angular momentum. However, since this
loss is proportional to $(R_1/a)^2$, (where $R_1$ is the 
neutron star's radius) it is negligible.

Combining together all the different mechanisms for loss of orbital
angular momentum, the total angular momentum loss rate is
\be
\frac{\dot{J}}{J} = \left(\frac{\dot{J}}{J}\right)_{GR}
+ \left(\frac{\dot{J}}{J}\right)_{\dot{M}}
+ \left(\frac{\dot{J}}{J}\right)_{\gamma}
+ \left(\frac{\dot{J}}{J}\right)_{B}.
\label{Jdot1}
\ee
Since the binary's orbital angular momentum is
\be
J = M_1 M_2 \left( \frac{Ga}{M_1+M_2} \right)^{1/2}
\ee
the change in orbital angular momentum due to rearrangements of the mass
is
\be
\left(\frac{\dot{J}}{J}\right)
= \frac12 \frac{\dot{a}}{a} + \frac{\dot{M_1}}{M_1}
 + \frac{\dot{M_2}}{M_2} - \frac12 \frac{(\dot{M_1} + \dot{M_2})}{(M_1+M_2)}.
\label{Jdot2}
\ee
Equations (\ref{adot}), (\ref{Rdot}), (\ref{Mdot}), (\ref{Jdot1}) and (\ref{Jdot2})
are a set of equations that allows the mass accretion rate
to be written in terms of the angular momentum loss rate due to 
various processes,
\be
\frac{\dot{M_2}}{M_2} = \frac32 \frac{1}{A(\alpha,\beta,\gamma,n_{ad},q)}
	\left[  \left(\frac{\dot{J}}{J}\right)_{GR}
	  +  \left(\frac{\dot{J}}{J}\right)_{B}
	  - \frac12 \left(\frac{\dot{R}_2}{R_2}\right)_{th} \right]
\ee
where the function $A(\alpha,\beta,\gamma, n_{ad},q)$ is defined by
\ba
A&=& \frac{5+3n_{ad}}{4}  - \frac32 q(1-\gamma)\beta
	-\frac12 \frac{q}{1+q}(1-\beta+\beta\gamma) \nonumber\\
&&	-\frac32 \alpha (1-\beta)(1+q)
	- \frac32 \delta \beta \gamma \frac{q^2}{1+q},
\label{eqA}
\ea
and it should be remembered that the adiabatic index $n_{ad}$
depends on $M_2 = qM_1$. 
When $\gamma = 0$, this equation reduces to the equation
derived by \citet{RJW82} for non-conservative mass transfer. 
In the binary systems of interest $q$ is small, so the term 
proportional to $\delta$ is explicitly very small. It should
also be remembered that the constant $\alpha$ is proportional
to $q^2$ if the mass loss occurs near to the neutron star,
which is the most likely situation.

If we were to ignore changes in the secondary due to 
thermal adjustments and any angular momentum losses due
to magnetic braking, a formula for the orbital period
derivative can be derived. (We will the discuss the
applicability of these assumptions in the next section.)
Making use of Kepler's law, we find the following equation
for the orbital period derivative
\be
\dot{P} = \frac{96 \pi}{5}  \frac{ q (1+q)^{-1/3}}{B(\alpha,\beta,\gamma,n_{ad},q)} 
\left( \frac{2 \pi GM_1}{ c^3 P} \right)^{5/3} ,
\ee
where the function $B(\alpha,\beta,\gamma,n_{ad},q)$ is defined by
\be
B = \frac{2}{1-3n_{ad}} A.
\ee
Making use of the mass function $f$,
\be
f = M_1 \frac{(q \sin i)^3}{ (1+q)^2},
\ee
the orbital period derivative is
\be
\dot{P} = \frac{96\pi}{5}     
\left( \frac{f (1+q)}{M_\odot}\right)^{1/3} \frac{1}{B \sin i}
\left( \frac{2 \pi GM_\odot}{ c^3 P} \right)^{5/3}
\left(\frac{M_1}{M_\odot}\right)^{4/3}
.  \label{Pdot} 
\ee
The predicted values of $\dot{P}$ can be written in the form
\be
\dot{P} = \frac{{\cal D}}{B} 
	\times 10^{-14} \frac{(1+q)^{1/3}}{\sin i}  
\left( \frac{M_1}{M_\odot}\right)^{4/3},
\ee
where the observationally determined constant ${\cal D}$ is defined by
\be
{\cal D} = 18.40 \times \frac{ f_{-6}^{1/3}} {P_3^{5/3}},
\label{CalD}
\ee
and $f/M_\odot = f_{-6} \times 10^{-6}$ and $P = P_{3} \times 10^3$ s.
Values for the constant ${\cal D}$ for the ultra-compact binaries
with measured mass functions
 are given in Table~\ref{table:1}.
Since the contribution to the period derivative from the relativistic
term $\gamma$ is quite small, we will typically require 3 significant
figures, hence the displayed accuracy in equation
(\ref{CalD}) and in Table \ref{table:1}. Although the binary mass 
ratio $q$ is quite small for the systems of interest we have
explicitly kept the factor of $(1+q)^{1/3}$ in the equation
for the period derivative since a precise result is required.
Present observations of the 
orbital period are not precise enough to detect
such a small period derivative in the orbit of any of the
systems listed in Table 1, but observations over a long period of time
may allow the detection of the period derivative.

A similar formula for the rate that mass is lost from the companion can 
be written in the form
\be
\dot{M}_2 = {\cal F} \times 10^{-12} \frac{(1+q)}{A (\sin i)^2} 
	\left(\frac{M_1}{M_\odot}\right)^2
	\frac{M_\odot}{yr},
\ee
where the constant ${\cal F}$ is defined by
\be
{\cal F} = 58.1 \times \frac{f_{-6}^{2/3}}{P_3^{8/3}}.
\label{CalF}
\ee
Values for the constant ${\cal F}$ are given in Table \ref{table:1}.
It has already been shown by \citet{CM98,BC01} that the mass transfer
rate for SAX J1808.4-3658 is consistent with being driven by gravitational
radiation. However, the determination of $\dot{M}_2$ is probably no
better than order of magnitude, so these measurements can't be thought
of as providing an extra constraint on $\gamma$.

\section{Astrophysical Considerations}
\label{s:evol}

The expression for the period derivative derived in the previous section
depends on the parameters $M_1, \sin i, n_{ad}, \alpha, \beta$, $\delta$  and $\gamma$.
(The mass ratio $q$ is not independent, since it is determined once
$f$ is measured and $M_1$ and $\sin i$ are specified.) This expression
was derived by assuming that the secondary's thermal timescale is very
long, and that magnetic braking is not operating. We now consider some
of the models for the evolution of ultra-compact binaries in order
to evaluate in which models there might be a chance of detecting
$\gamma$. 

There are two main scenarios for producing a neutron star 
X-ray binary with a very short orbital period (see \citet{NR03}
for a discussion). In the first scenario, the companion star is
a main sequence star and will be denoted MS-NS. The alternative 
picture is of a white dwarf companion and will be denoted WD-NS.

In the MS-NS scenario, mass transfer from a low mass star occurs 
near the time when Hydrogen is depleted in its core
 \citep{NR03}. Both gravitational
radiation and magnetic braking \citep{EA99} remove orbital angular momentum
from the system, and the orbital periods can evolve down to very short timescales
\citep{PRP02}. In this type of system, magnetic braking could easily be as important
as gravitational radiation in the removal of angular momentum. Since the parameters
describing magnetic braking are not accurately known, it would not be possible to 
measure the effect of the $\gamma$ term in this type of binary system.

In the WD-NS scenario the white dwarf fills its Roche lobe when the orbital period
is very small (3 minutes for example) and gravitational radiation drives mass transfer and
increases the orbital period to the range of 40 minutes, reducing
the white dwarf mass to very low values of order $0.01 M_\odot$ \cite{RPR00}. 
Magnetic braking is unlikely to be operating in the WD-NS picture, since it is 
most effective if the donor star has a large radius,
and it typically assumes a convective envelope that drive a magnetic dynamo.
A more difficult question
is whether the white dwarf is being heated, through either being irradiated by
the X-ray flux from the neutron star or through tidal effects. If the heating timescale is short
compared to the mass-loss timescale, the $(\dot{R}_2/R_2)_{th}$ term can't be
discarded. \citet{DB03} have shown that observations of $\dot{M}_2$ and $P$ for WD-NS systems 
can constrain possible combinations of WD composition and temperature and have demonstrated
that it is possible to use these observations to show that in some cases an adiabatic evolution is
consistent with observations. 

If a measurement of the orbital period derivative of a WD-NS binary were to be made,
and if one assumed that the evolution is adiabatic, equation (\ref{Pdot}) could be used
to constrain a combination of $M_1, \sin i, n_{ad}, \alpha, \beta, \delta$ and $\gamma$.
In the compact systems most likely of interest, $q$ is probably quite small, of 
order $0.01$, so terms of order $q^2$ in the function $A$ (such as
the term proportional to $\delta$) can be 
neglected. If any mass loss take place near the neutron star (say through a propeller 
mechanism) then the parameter $\alpha \sim q^2$ and the $\alpha$ term can be neglected.
In the function $A$, $\gamma$ always appears in the combination
$q\beta(1-\gamma)$, so it is essentially this term that we would like to measure.  
Since $\gamma$ could realistically be expected to have a value near $0.2$ if we
were pessimistic and supposed that $\beta$ could be as small as $0.5$, the 
term $q\beta(1-\gamma) \ge 4\times 10^{-3}$. We would then want to know $n_{ad}$ to
an accuracy of at least $\pm 0.001$ in order to have a hope of 
measuring the combination of $\beta(1-\gamma)$, which would give an upper limit on
the value of $\gamma$.

\begin{figure*}[t]
\epsscale{0.5}
\plotone{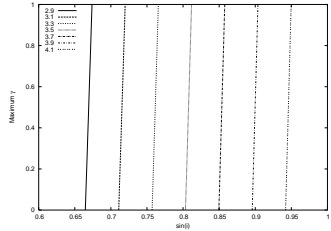}
\caption{Plots of the maximum allowed values of $\gamma$ versus binary
inclination angle for XTE~J1751-305, assuming a Helium white dwarf companion
with $M_2 = 0.02 M_\odot$ and an adiabatic index of $n_{ad}=-0.22$. In this
plot, each curve corresponds to a value of the orbital period derivative
in units of $10^{-14} s/s$. 
\label{fig3}}
\end{figure*}

If the WD temperature and composition 
were known then $n_{ad}$ is given by \cite{DB03} as a function of $M_1$ and 
$\sin i$. For a given WD temperature and composition, a period derivative 
measurement (along with all of the other assumptions listed in this section)
would yield a constraint on a combination of $M_1, \sin i,$ and $\beta(1-\gamma)$. 
Further constraints on the inclination angle would strengthen this constraint.
For example recent observations 
of SAX J1808.4-3658 by \cite{Wang01}
have made use of a reddening distance technique and models of
X-ray heated accretion disks to constrain the 
binary inclination, depending on the assumed distance to the binary system.

In the equations for the period derivative for mass transfer induced only by 
gravitational radiation with adiabatic changes to the white dwarf, the value
of the $n_{ad}$ plays the most important role (along with the inclination
angle and the neutron star mass) in determining the binary's evolution. Ideally,
observations of such as system ignoring the relativistic $\gamma$ term could
be used to constrain ranges of values of $n_{ad}$ consistent with an 
adiabatic evolution for various possible values of $\sin i$ 
and $M_1$.  Once these values are determined, and expanding equation (\ref{eqA})
to lowest order in $q$ (assuming that the mass loss occurs near the neutron star)
one finds the limit on $\gamma$,
\ba
\gamma &\le& \frac32 - \frac{1}{q} \left(
\frac{(5+3n_{ad})}{4} \right. \nonumber \\
&& \left.
- \frac{(1-3n_{ad})}{2} \frac{{\cal{D}}\times 10^{-14}}{\dot{P}}
\frac{(1+q)^{1/3}}{\sin i}  
\left( \frac{M_1}{M_\odot}\right)^{4/3}\right)
\label{limit}
\ea
where we have used the fact that $\beta \le 1$. Although a number of 
assumptions have been made in deriving this limit, inequality (\ref{limit})
does have the potential to provide an interesting limit on $\gamma$. 
Since the mass ratio $q$ is expected to be small (of order 0.01), a meaningful
limit will only occur if the terms proportional to $1/q$ are similar in magnitude,
so that their difference is of order $q$. Since $\cal{D}$ is of order unity,
only orbital
period derivatives of order $10^{-14} s/s$ will provide an interesting limit on $\gamma$,
given all of our assumptions. From equation (\ref{limit}) it can be seen that if $\dot{P}$ is too
small, the maximum allowed value of $\gamma$ will be so large as to be an uninteresting limit.
If however, $\dot{P}$ is too large, the maximum allowed value of $\gamma$ will be negative,
implying that our assumptions about the physics of the binary are invalid.

As an example, consider the situation where the companion star is a Helium white dwarf
with mass $0.02 M_\odot$ and an adiabatic index of $n_{ad}=-0.22$, consistent with 
a temperature near $10^7 K$ as calculated in the models of \cite{DB03}. With data for the mass
function for 
XTE J1751-305, inclination angles with $\sin i \ge 0.7$ give neutron star masses larger than 
1.4 $M_\odot$. Given this range of inclination angles, it is easy to use equation (\ref{limit})
to find the maximum allowed value of $\gamma$ for any observed $\dot{P}$. These values are
shown in Figure 3. This figure can be interpreted by noting that if the binary's inclination
angle were found to be $\sin i = 0.9$, then a value of $\dot{P}= 3.9 \times 10^{-14} s/s$
would be consistent with our assumptions, while a larger period derivative (say $4.1 \times 10^{-14}$)
would yield a negative $\gamma$ and is inconsistent with our assumptions.

\section{Conclusions}
\label{s:conclusions}

In this paper we have discussed the effect on the orbital period
of changing a neutron star's binding energy through accretion.
 We have shown that it is possible,
in principle, for an observation of an orbital period derivative
and with some knowledge of the donor star to place constraints
on possible values of the neutron star's mass and the 
binding energy parameter that depend on the binary's inclination angle.
However, there 
are a number of uncertainties inherent in our method. We will
now discuss these potential problems.

Our calculation assumes that the binary's orbit is given by
newtonian physics and have not considered post-newtonian (PN)
corrections. However, the 1st PN corrections
enter at the level of $v^2/c^2$ where $v$ is the orbital velocity.
Since we are considering wide (from a relativist's point of view)
binaries with separations of the order of $10^4$ km, the 1st PN 
corrections will be about 1000 times smaller than the orbital
corrections due to the change in binding energy discussed in this 
paper.

The model of mass transfer considered in this paper 
makes use of the Roche model, ignores
the spin angular momentum of the secondary, and calculates
the angular momentum loss by gravitational radiation in the
point particle approximation. The validity of these
assumptions has been examined by \cite{RUY01} for the
case of a white dwarf primary. \cite{RUY01} do an
accurate numerical integration to compute the structure 
of a semi-detached compressible co-rotating binary in hydrostatic
equilibrium. They find that Paczynski's approximation
for the Roche lobe radius and the expression for the
total angular momentum are correct to about 1\% if
the secondary star's spin angular momentum is included.
In their computations, they also compute the rate that
angular momentum is lost when the secondary's structure
is taken into account and find that the point-particle
approximation introduces an error of about 1\%. 
While these corrections are quite small, the relativistic
corrections due to the neutron star's structure are also
quite small. For this reason it would be desirable to 
consider more realistic models such as those computed
by \cite{RUY01}.

Larger uncertainties lie in the assumed physics of the 
binary system. We need to restrict our attention to systems where
the donor star is a white dwarf, and the angular momentum loss is
strictly through gravitational radiation. It is also important that
the donor star is not rapidly heated so that it is possible to
assume that the donor's radius responds adiabatically to mass loss.

If any mass is
lost from the binary system, the angular momentum lost 
could potentially drive a period derivative correction
which is larger than the relativistic effect considered here.
If the mass loss occurs near the neutron star, (through
the propeller mechanism, for example) the effect is not
large enough to be of importance for very low mass companions. 
If mass is lost from a region near the companion, then
the large angular momentum loss would make it impossible to
measure the relativistic effect. 

The gravitational-radiation-driven angular momentum loss discussed in this paper is
a continuous process, while the mass transfer onto the star is highly episodic 
in that matter can be stored in the accretion disk for years before being transfered
onto the neutron star. Since our treatment assumes a continuous transfer of mass,
measurements of the orbital period derivative would have to be measured over the
course of many years before there could be any hope of measuring the 
relativistic parameter $\gamma$.

The orbital frequencies typical for the binaries of interest lie
within the bandwidth of the Laser Interferometer Space Antenna (LISA).
The dimensionless gravitational-wave amplitude $h$ \citep{Tho98}
measured by LISA can be written as 
\be
h \le \frac{10^{-23}}{2^{4/3}} f_{-6}^{1/3} \left(\frac{M_1}{2M_\odot}\right)^{4/3}
	\left( \frac{2 \hbox{hr}}{P}\right)^{2/3} \left( \frac{2 \hbox{kpc}}{D}\right)
\ee
where $D$ is the distance to the binary.
Using the measured binary parameters for SAX J1808.4-3658,
we find that the gravitational-wave amplitude is $h\le 10^{-23}$,
which is well below the ``confusion noise'' produced by the 
population of galactic white dwarfs \citep{Nel01}. 
The low amplitude of the signal
is mainly due to the small mass ratio, so we expect that the
binaries of the type discussed in this paper will not be detectable
by LISA, unless new data analysis techniques are developed which
can take advantage of the known positions of the binaries.

In order for the effects of a changing relativistic mass
deficit to be observable, very small changes in the binary's
orbital period must be measured. In addition, the rate of
mass transfer must be known in order to calculate the
magnitude of the relativistic effect. For this reason,
we have examined the scenario of mass transfer from a 
fully degenerate dwarf driven
by gravitational radiation, since the mass transfer rate
is related quite simply to the binary's parameters. We have
focused our attention on the ms spin-period neutron
stars since the prospects for measuring the orbital 
period for these systems seems most promising. However, the
orbital effect of a changing mass deficit also holds for non-rotating
stars. In fact, the effects are much simpler in the case of 
a non-rotating star, since the effects of torquing due to
mass accretion are completely negligible.

\acknowledgments

This research was supported by a grant from NSERC. We would like
to thank Saul Rappaport for his very useful comments on the 
physics in this paper. We also thank Symeon Konstantinidis
for pointing out an error in a previous version of this paper.

\begin{deluxetable}{rlllllll}
\tablewidth{0pc}
\tablecolumns{8}
\tablecaption{
Observational data 
for accreting ms pulsar systems and values of the constants ${\cal D}$ and ${\cal F}$.
\label{table:1}}
\tablehead{
\colhead{Binary System}
&\colhead{$\nu_s$ (Hz)\tablenotemark{a}} &\colhead{$\dot{\nu}_s (\hbox{Hz}\, \hbox{s}^{-1})$}
&\colhead{$P$ (s)\tablenotemark{b}}& \colhead{$f \; (M_\odot)$\tablenotemark{c}}
&\colhead{${\cal D}\tablenotemark{d}$}&\colhead{${\cal F}\tablenotemark{e}$}
&\colhead{Reference}}
\startdata
SAX J1808.4-3658 & 401 & $< 7\times 10^{-13}$ & 7249.119 & $3.7789 \times 10^{-5}$ 
& 2.274 &3.3 &\citet{CM98}\\
XTE J1751-305    & 435 & $< 3 \times 10^{-13}$& 2545.3414& $1.2797 \times 10^{-6}$ 
& 4.210 &5.7 &\citet{Markwardt02}  \\
XTE J0929-314    & 185 & $- 9.2 \times 10^{-14}$ & 2614.746 & $2.7 \times 10^{-7}$ 
&    1.1 &0.4 &\citet{Galloway02}\\
XTE J1807-294    & 191 &  & 2404.2& & & & \citet{Mark03}\\
XTE J1814-338    & 314 &  & 15390 & & & & \citet{Stroh03}\\
\enddata 
\tablenotetext{a}{Pulsar spin frequency}
\tablenotetext{b}{Orbital Period}
\tablenotetext{c}{Mass function}
\tablenotetext{d}{$\cal{D}$ is defined in equation (\ref{CalD}).}
\tablenotetext{e}{$\cal{F}$ is defined in equation (\ref{CalF}).}

\end{deluxetable}

\end{document}